\documentclass[runningheads]{llncs}
\usepackage{amsfonts,amssymb}
\usepackage{amsmath}
\usepackage{float}
\usepackage[T1]{fontenc}
\usepackage{pifont}
\usepackage{multirow}
\usepackage{graphicx}
\usepackage[table,xcdraw]{xcolor}
\usepackage{graphicx}
\usepackage{color}
\usepackage{hyperref}
\hypersetup{colorlinks = True, 
	    linkcolor = red, 
        citecolor = blue} 
        
\usepackage[misc]{ifsym}

\begin{document}
\title{M\textsuperscript{4}oE: A Foundation Model for Medical Multimodal Image Segmentation with Mixture of Experts}
\author{Yufeng Jiang\inst{1}\orcidID{0009-0004-4987-2683} \and
Yiqing Shen\inst{2}\textsuperscript{(\Letter)}\orcidID{0000-0001-7866-3339}}
\authorrunning{Y. Jiang and Y. Shen}
\institute{Department of Computer Science, Hong Kong Baptist University, Hong Kong SAR, China
\and
Department of Computer Science, Johns Hopkins University, Baltimore, USA
\email{yshen92@jhu.edu}
% \\
}
\maketitle             
\begin{abstract}
Medical imaging data is inherently heterogeneous across different modalities and clinical centers, posing unique challenges for developing generalizable foundation models.
Conventional entails training distinct models per dataset or using a shared encoder with modality-specific decoders. 
However, these approaches incur heavy computational overheads and suffer from poor scalability.
To address these limitations, we propose the Medical Multimodal Mixture of Experts (M\textsuperscript{4}oE) framework, leveraging the SwinUNet architecture.
Specifically, M\textsuperscript{4}oE comprises modality-specific experts; each separately initialized to learn features encoding domain knowledge. 
Subsequently, a gating network is integrated during fine-tuning to modulate each expert's contribution to the collective predictions dynamically. 
This enhances model interpretability and generalization ability while retaining expertise specialization.
Simultaneously, the M\textsuperscript{4}oE architecture amplifies the model's parallel processing capabilities, and it also ensures the model's adaptation to new modalities with ease.
Experiments across three modalities reveal that M\textsuperscript{4}oE can achieve 3.45\% over STU-Net-L, 5.11\% over MED3D, and 11.93\% over SAM-Med2D across the MICCAI FLARE22, AMOS2022, and ATLAS2023 datasets.
Moreover, M\textsuperscript{4}oE showcases a significant reduction in training duration with 7 hours less while maintaining a parameter count that is only 30\% of its compared methods.
The code is available at \url{https://github.com/JefferyJiang-YF/M4oE}.

\end{abstract}

\section{Introduction}
\begin{figure*}[t!]
    \centering
    \includegraphics[width=1\linewidth]{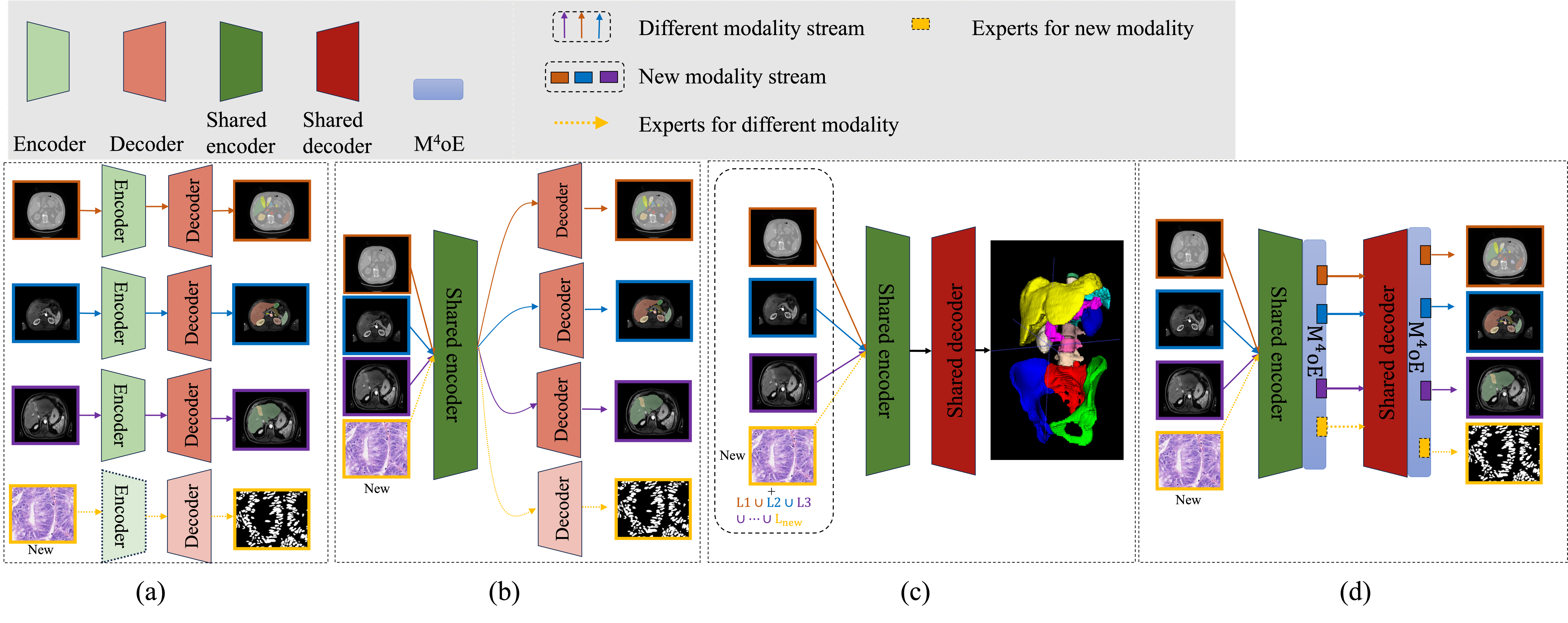}
    \caption{Paradigms for multimodal medical segmentation: (a) One network for each modality; 
    (b) A shared encoder with modality-specific decoders;
    (c) Union of labels with pseudo-label training;
    (d) Our proposed M\textsuperscript{4}oE with shared encoder and decoder.
    }
    \label{fig:different_paradiam}
\end{figure*}
The advent of vision foundation models such as Segment Anything Model (SAM) and Stable Diffusion \cite{rombach2022high} has catalyzed remarkable progress in natural image understanding~\cite{liu2023towards}. 
However, translating these advancements to medical imaging poses notable challenges \cite{10.36227/techrxiv.19103432.v1} due to the heterogeneity across modalities like MRI, CT, and X-rays, each exhibiting distinct data characteristics \cite{huang2020unet3+}. 
Consequently, medical imaging necessitates specialized processing beyond the capabilities of models designed for more homogeneous natural images.
Historically, coping with multimodal medical data heterogeneity entailed developing customized models targeting individual datasets, as depicted in Fig.~\ref{fig:different_paradiam}(a).
Although minimizing initial complexity, this approach introduces substantial overheads for adapting models to new datasets. 
Alternatively, following the multi-task learning design, some methods utilize a shared encoder with modality-specific decoders to extract common representations while retaining tailored processing~\cite{chen2019med3d,zhang2021dodnet}, as in Fig.~\ref{fig:different_paradiam}(b).
However, computational requirements can still scale poorly as data diversity grows \cite{johnson2019mimic}.
Unified encoder-decoder schemes also exist to standardize label processing ~\cite{liu2022universal_unfiying_label}, as shown in Fig.~\ref{fig:different_paradiam}(c), but may compromise extensibility to new modalities \cite{anwar2018medical/review}.

To address these shortcomings, we introduce Medical Multimodal Mixture of Experts (M\textsuperscript{4}oE), built upon SwinUNet \cite{cao2022swinunet}, to effectively process multimodal medical images segmentation.
The conventional Transformer architecture struggles to adapt to modality-specific attributes optimally \cite{shin2021perspectives_crossmodal}.
To overcome this gap, M\textsuperscript{4}oE initializes distinct experts focusing on unique aspects of each modality (Fig. \ref{fig:different_paradiam}(d)).  
A gating network then dynamically combines expert outputs during inference. 
The design of M\textsuperscript{4}oE enhances parallel processing capabilities and simplifies adapting to new modalities without extensive reconfiguration.

The contribution of this work is three-fold, summarized as follows.
1) We propose an innovative foundation model for multimodal medical image segmentation based upon a mixture of experts, named M\textsuperscript{4}oE.
It allows efficient scaling to large heterogeneous datasets without drastic parameter increases, striking an optimal balance between broad applicability and computational efficiency.
2)We propose M\textsuperscript{4}oE's two-phase training with modality-specific experts and, a fusing gating network, and a shuffle-adaptable linear projection architecture for multi-modality and label mapping.
%2) We propose a novel two-phase training strategy specifically for M\textsuperscript{4}oE where we first train experts to learn modality-specific, and then a gating network learns to fuse them.
%
%3) We designed a linear mapping architecture with shuffle, adaptable to multi-modality and labels, auto-adjusting output channels per class vis predictive head.
%
3) Our comprehensive experimental evaluations indicate that M\textsuperscript{4}oE achieves competitive performance on multimodal medical datasets and shows promising transferability, with reduced need for reconfiguration across different modalities.
\begin{figure}[t]
    \centering
    \includegraphics[width=1\linewidth]{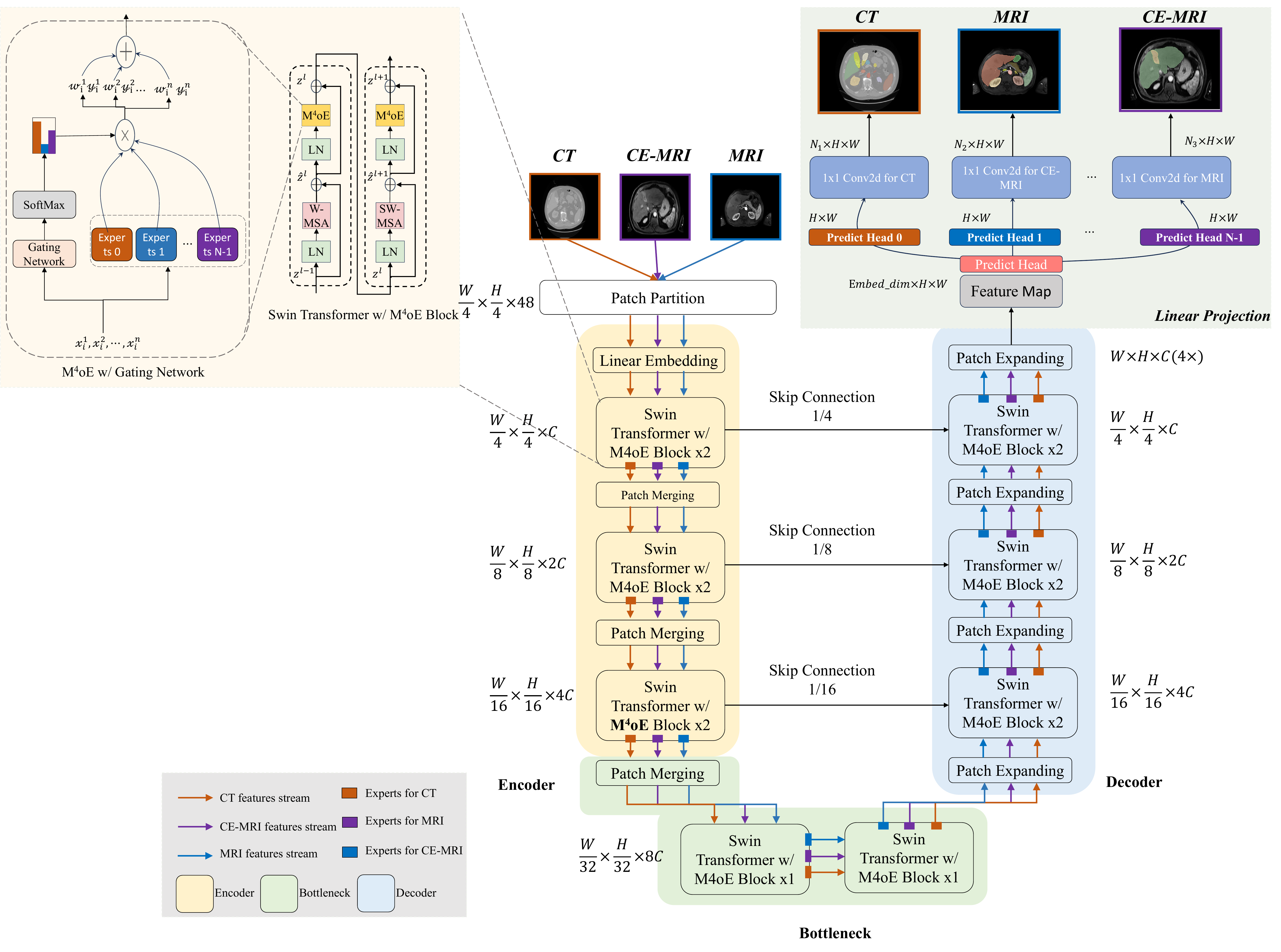}
    \caption{Overall framework of M\textsuperscript{4}oE implemented on a SwinUNet backbone.
    At the core is the M\textsuperscript{4}oE with a gating network, which dynamically selects specialized experts for different imaging modalities (CT, CE-MRI, MRI); each expert learns representation for its respective data characteristics.
    }
    \label{fig:M4oESwinUNET}
\end{figure}

\section{Methods}
We present M\textsuperscript{4}oE building on SwinUNet \cite{cao2022swinunet}, where our model substitutes the MLP component with a Mixture-of-Experts designed specifically for multimodal medical image segmentation to improve representation.

\subsubsection{Swin Transformer Backbone} 
The core backbone of our architecture leverages Swin Transformer blocks \cite{liu2021swin}. 
Contrasting conventional multi-headed self-attention (MHSA), Swin Transformers employ a local, sliding-window variant termed window-based MHSA (W-MSA).
As illustrated in Fig. \ref{fig:M4oESwinUNET}, each Swin Transformer block comprises a LayerNorm, a Multilayer Perceptron (MLP), W-MSA, and a shifted-window MHSA (SW-MSA).
The sliding window approach limits self-attention computation to local patch regions, reducing complexity compared to global MHSA. 
% 
% Combined with shifted window partitioning, this allows the network to capture both local features and cross-region dependencies.

\subsubsection{Medical Multi-Modal Mixture-of-Experts (M\textsuperscript{4}oE) Formulation}
Our key innovation involves substituting the MLP layer in each Swin Transformer block with a M\textsuperscript{4}oE, as depicted in Fig. \ref{fig:different_paradiam}(d). 
This comprises a gating network along with a set of modality-specific expert networks.
The experts themselves consist of two-layer MLPs, utilizing GELU for non-linear activation.
Each expert network specializes in learning representations corresponding to a particular imaging modality.
This allows dynamically allocating computational resources conditional on data complexity, enhancing the model's representation capacity.
The gating network leverages a simple single-layer MLP activated by a softmax function to weigh the contribution of each expert dynamically based on the input.

\subsubsection{M\textsuperscript{4}oE in Swin Transformer}
We incorporate the proposed M\textsuperscript{4}oE into Swin Transformer blocks as follows:
\begin{equation}
\begin{aligned}
\Tilde{z}^l &= \text{W-MSA}(\text{LN}(z^{l-1})) + z^{l-1}, \\
z^l &= \text{M}^{\text{4}}\text{oE}(\text{LN}(\Tilde{z}^l)) + \Tilde{z}^l, \\
\Tilde{z}^{l+1} &= \text{SW-MSA}(\text{LN}(z^l)) + z^l, \\
z^{l+1} &= \text{M}^{\text{4}}\text{oE}(\text{LN}(\Tilde{z}^{l+1})) + \Tilde{z}^{l+1},
\end{aligned}
\end{equation}
where $\Tilde{z}$ and $z$ denote the W/SW-MSA and M\textsuperscript{4}oE representations of the $l^{th}$ block respectively.
Within M\textsuperscript{4}oE, the gating network assigns weights to each expert's output as follows:
\begin{equation}
\begin{aligned}
    \text{weight}_G & = \text{SoftMax} (W_g \cdot \mathbf{x} + \mathbf{b}_g) = \frac{\exp(\mathrm{weights_G})}{\sum_j\exp(\mathrm{weights_G}_j)},\\
\text{where}\quad &\mathrm{expert}_i(\mathbf{x})=W_{o,i}\cdot\mathrm{GELU}(W_{h,i}\cdot\mathbf{x}+\mathbf{b}_{h,i})+\mathbf{b}_{o,i},
\end{aligned}
\end{equation}
where $\text{weight}_G$ is the gating weights to the input $\mathbf{x}$, the gating weights are computed using the SoftMax function for assigning different weights based on other representations. 
$W_{h,i},\mathbf{b}_{h,i},W_{o,i},\text{and }\mathbf{b}_{o,i}$ are the learnable weights and biases of the i\textsuperscript{th} expert network. 
The outputs of all experts are computed and stacked together to form a tensor \textit{i}.\textit{e}., $\text{expert}\_1(\mathbf{x}), \text{expert}\_2(\mathbf{x})  \dots \text{expert}\_{\text{num\_experts}}(\mathbf{x})$. 
It follows that the output of the M\textsuperscript{4}oE module is the weighted sum of all expert outputs, namely
\begin{equation}
\mathrm{outputs}=\sum_{i=1}^\text{n}{\text{weight}_i}\cdot\mathrm{expert}_i(\mathbf{x}),
\end{equation}
where $\text{weight}_{i}$ is the i\textsuperscript{th} element of the weights vector, representing the weight assigned to the i\textsuperscript{th} expert's output.

\subsubsection{Dynamic Selection for Linear Projection Head}
A core challenge in multimodal segmentation involves handling varying class numbers across modalities. 
We introduce a dynamic selection mechanism through customizable linear projection heads inserted before output to address this.
First, feature maps are padded to equalize dimensions across all samples regardless of modality.
Subsequently, each sample's linear projection layer dynamically selects only the logits corresponding to the classes present.
Thereby, the final output dimensions are aligned with the native label distribution in a modular, input-conditional manner.
This process grants invariance to shuffling across modalities during training. 
Meanwhile, the lightweight projections minimize overheads, allowing efficient reconfiguration of new datasets.

\subsubsection{Two-phase Training Strategy}
We propose a novel two-phase training strategy consisting of expert pre-training and gate network fine-tuning phases. 
In the expert pre-training phase, we leverage the Masked Autoencoder(MAE) \cite{he2022maskedMAE} to pre-train each expert. This enables it to learn meaningful representations for its modality without the gating network.
Subsequently, in the gate network fine-tuning phase, we load the pre-trained encoder and expert parameters and incorporate the gating network. 
We then fine-tune the entire M\textsuperscript{4}oE model end-to-end on downstream segmentation tasks.
This two-phase approach allows the experts to specialize in their respective modalities before learning to selectively combine their representations via the gating network.
\begin{table}[!t]
\centering
\caption{
% Comparing methods in terms of pre-training, dimensionality, multimodality, parameter count, training hours, and the parameters added when introducing new modalities.
Comparison of M\textsuperscript{4}oE against prior arts in terms of crucial model design choices and computational requirements, including pre-training strategy, input dimensionality, multimodality handling capability, number of parameters, training duration, and parameters added when introducing new modalities.
}
\label{tab:method_compare}
\resizebox{\textwidth}{!}{%
\begin{tabular}{c|ccc|ccc}
\hline
Methods                     & Pre-train              & 2D/3D & Multimodaliy                & \#Param & Training hours & Introduce new modality \\ \hline
\multicolumn{1}{c|}{DoDNet~\cite{zhang2021dodnet}} & \textcolor{red}{\ding{55}}                 & 3D    & \multicolumn{1}{c|}{\textcolor{green}{\ding{51}}} & 17.3M   & 48 hours       & -                \\
\multicolumn{1}{c|}{MED3D~\cite{chen2019med3d}}  & \textcolor{red}{\ding{55}}                 & 3D    & \multicolumn{1}{c|}{\textcolor{green}{\ding{51}}} & 117.51M & 16 hours       & 21M              \\
\multicolumn{1}{c|}{Liu's~\cite{liu2022combining/FLARE/compare}}  & \textcolor{red}{\ding{55}}                 & 3D    & \multicolumn{1}{c|}{\textcolor{red}{\ding{55}}} & 6.66M   & 19.5 hours     & 6.66M                \\
\multicolumn{1}{c|}{STU-Net-L~\cite{huang2023stu}} & \textcolor{green}{\ding{51}} Supervised & 3D & \multicolumn{1}{c|}{\textcolor{red}{\ding{55}}} & 440.30M & 156.8 hours & 440.30M \\ \hline
\multicolumn{1}{c|}{Ours}                        & \textcolor{green}{\ding{51}} Self-supervised & 2D    & \multicolumn{1}{c|}{\textcolor{green}{\ding{51}}} & 27.31M  & 12 hours       & 10M        \\ \hline
\end{tabular}%
}
\end{table}

\section{Experiments}
\subsubsection{Implementations}
All experiments leveraged a single RTX 4090 GPU system with 24GB memory, running PyTorch 1.11.0, Python 3.8 on Ubuntu 20.04, and CUDA 11.3.
To initialize feature representations, the M\textsuperscript{4}oE in encoder was pre-trained via MAE \cite{dai2023swinMAE,he2022maskedMAE}.
The decoder, optimization settings, input resolution, and patch size were configured identically to SwinUNet \cite{cao2022swinunet} for fair comparison. 
Specifically, training employed a batch size of $36$ and weight decay of $10^{-4}$ using AdamW optimization.

\subsubsection{Datasets}
We leverage three multimodal medical image segmentation datasets to evaluate M\textsuperscript{4}oE's performance, namely FLARE22 \cite{FLARE22}, AMOS22 \cite{DBLP:conf/nips/JiBGYZZLZMW022/AMOS}, and ATLAS23 \cite{quinton2023tumour/ATLAS}.
Specifically, from FLARE22, we utilize 50 labeled CT scan cases.
AMOS22 provides 600 labeled cases comprising 500 CT and 100 MRI abdominal scans.
Additionally, we incorporate 60 cases of contrast-enhanced MRI (CE-MRI) liver scans with labels from ATLAS23.
We follow the previous work~\cite{huang2023stu} for data split.
While FLARE22 and AMOS22 focus on segmenting abdominal organs, ATLAS23 specializes in segmenting liver and tumors. 
This allows assessing M\textsuperscript{4}oE's capability in handling multimodal datasets across diverse anatomical segmentation tasks.

\begin{figure}[!t]
    \centering
    \includegraphics[width=0.9\linewidth]{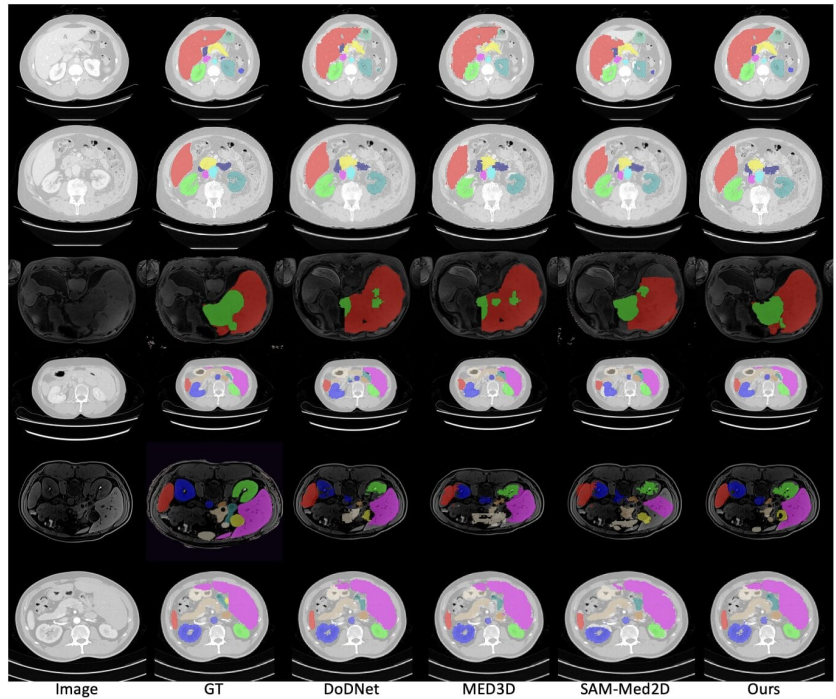}
    \caption{Visualization of segmentation results from various methods.
    The first and second rows depict FLARE22 with CT images, the third row showcases ATLAS2023 with CE-MRI images, the fourth and sixth rows display AMOS2022 with CT images, and the fifth row presents MRI images from AMOS2022. 
    The six columns from left to right correspond to the original image, the ground truth (GT), the DoDNet results, the MED3D results, the SAM-MED2D results, and our M\textsuperscript{4}oE results.}
    \label{fig:enter-label}
\end{figure}

\subsubsection{Results}
As shown in Table~\ref{tab:method_compare}, DoDNet~\cite{zhang2021dodnet}, MED3D~\cite{chen2019med3d}, and Liu's~\cite{liu2022combining/FLARE/compare} methods lack pre-training, potentially limiting learned representations.  
Although focusing on 3D data, DoDNet and MED3D partially mitigate this via multimodality. 
MED3D's efficiency is evident with its short training duration despite a high parameter count, while Liu's method, with fewer parameters, might be better generalized. 
Meanwhile, STU-Net-L's~\cite{huang2023stu} supervised pre-training potentially improves initial learning but necessitates labeled data, and its high parameter count and extended training time could be drawbacks.
Our approach introduces self-supervised pre-training on 2D data, eliminating the need for labeled datasets and showing cost-effectiveness. 
It balances model complexity and resource efficiency, proving advantageous for practical applications where performance and efficiency are paramount.
As shown in Table~\ref{tab:dsc_results}, our approach excels in the FLARE22 and AMOS-CT datasets, with the highest mean dice scores (DSC), indicating its suitability for multimodal tasks and accurate segmentation. 
Our method shows slight performance decreases when introducing MRI modality, evidencing its adaptability, whereas DoDNet and MED3D exhibit notable performance drops.
Our model's training across diverse datasets likely contributes to its robustness, as opposed to STU-Net-L, which although achieving high DSC in a single-task setting, lacks versatility.
SAM-Med2D's lower performance might stem from its 2D focus, potentially failing to capture 3D medical imaging complexities.
Our method outperforms the ATLAS2023 dataset, maintaining robustness with new modalities and leading in DSC and mIoU metrics, as seen in Table~\ref{tab:atlas-result}. 
This underscores the limitations of DoDNet and MED3D in modality variations, while our model's balanced approach offers a promising direction for multimodal medical image segmentation.

\begin{table}[!t]
\centering
\caption{Comparison of DSC (\%) across multimodal segmentation methods. Our M\textsuperscript{4}oE model outperforms others, achieving the highest DSC in FALRE and AMOS-CT tasks. It shows adaptability and improvement (*) upon introducing a new modality, contrary to DoDNet and MED3D, which see performance drops. Unlike STU-Net-L's single-dataset focus, our model's robustness is attributed to training on three varied datasets. SAM-Med2D's performance is evaluated using a single data point. Results are highlighted in {\color[HTML]{FE0000}red} for best and {\color[HTML]{3531FF} blue} for second best.}
\label{tab:dsc_results}
\resizebox{\textwidth}{!}{%
\begin{tabular}{c|c|cccccccc}
\hline
Datasets                  & Method    & Liver & Left Kidney & Right Kidney & Gallbladder & Pancreas & Spleen & Stomach & Mean                         \\ \hline
                          & DoDNet~\cite{zhang2021dodnet}    & {\color[HTML]{3531FF} 96.64} & {\color[HTML]{FE0000} 95.43}       & {\color[HTML]{FE0000} 95.32}       & 74.58       & {\color[HTML]{3531FF} 79.48}    & 90.91  & 90.85   & {\color[HTML]{3531FF} 89.03} \\
                          & MED3D~\cite{chen2019med3d}     & 90.36 & 87.23       & 88.47        & 67.84       & 73.82    & {\color[HTML]{3531FF} 91.98}  & {\color[HTML]{FE0000} 92.30}   & 84.53                        \\
                          & Liu's~\cite{liu2022combining/FLARE/compare}     & 89.30  & 76.32       & 78.40         & 55.35       & 58.86    & 83.07  & 67.82   & 72.73                        \\
                          & STU-Net-L~\cite{huang2023stu} & 94.48 & 87.51       & 86.68        & {\color[HTML]{3531FF} 76.31}       & {\color[HTML]{FE0000} 82.00}    & 91.27  & 89.31   & 86.79                        \\
                          & SAM-Med2D~\cite{cheng2023sammed2d} & 80.37 & 77.96       & 78.19        & 50.37       & 53.12    & 82.47  & 75.01   & 71.07                        \\
\multirow{-6}{*}{FLARE22} & Ours      & {\color[HTML]{FE0000} 97.08} & {\color[HTML]{3531FF} 94.62}       & {\color[HTML]{3531FF} 94.92}        & {\color[HTML]{FE0000} 79.38}       & 76.99    & {\color[HTML]{FE0000} 95.74}  & {\color[HTML]{3531FF} 89.78}   & {\color[HTML]{FE0000} 89.78} \\ \hline
                          & DoDNet~\cite{zhang2021dodnet}    & 92.14 & 83.93       & 82.48        & 70.73       & {\color[HTML]{3531FF} 73.69}    & 90.58  & {\color[HTML]{3531FF} 85.14}   & 82.67                        \\
                          & MED3D~\cite{chen2019med3d}     & 88.25 & 85.11       & 85.29        & 65.12       & 59.78    & 88.37  & 81.10   & 79.00                        \\
                          & STU-Net-L~\cite{huang2023stu} & {\color[HTML]{3531FF} 95.47} & {\color[HTML]{3531FF} 88.37}       & {\color[HTML]{FE0000} 89.74}        & {\color[HTML]{FE0000} 76.92}       & {\color[HTML]{FE0000} 82.64}    & {\color[HTML]{FE0000} 93.24}  & {\color[HTML]{FE0000} 87.03}   & {\color[HTML]{FE0000} 87.63} \\
                          & SAM-Med2D~\cite{cheng2023sammed2d}  & 67.51 & 71.09       & 68.31        & 57.93       & 51.24    & 76.12  & 72.93   & 66.45                        \\
\multirow{-5}{*}{AMOS-CT} & Ours      & {\color[HTML]{FE0000} 95.59} & {\color[HTML]{FE0000} 88.72}       & {\color[HTML]{3531FF} 88.31}        & {\color[HTML]{3531FF} 73.94}       & 61.39    & {\color[HTML]{3531FF} 90.59}  & 82.78   & {\color[HTML]{3531FF} 83.04} \\ \hline
                          & DoDNet~\cite{zhang2021dodnet}    & 89.31 & 80.33       & 82.89        & 70.13       & 68.32    & 85.80  & 83.23   & 80.01                        \\
                          & MED3D~\cite{chen2019med3d}     & 83.42 & 82.11       & 86.33        & 69.12       & 63.20    & 86.23  & 80.79   & 78.74                        \\
                          & STU-Net-L~\cite{huang2023stu} & {\color[HTML]{FE0000} 95.59} & {\color[HTML]{3531FF} 89.00}       & {\color[HTML]{FE0000} 90.10}        & {\color[HTML]{FE0000} 79.97}       & {\color[HTML]{FE0000} 94.08}    & {\color[HTML]{FE0000} 94.37}  & {\color[HTML]{FE0000} 94.42}   & {\color[HTML]{FE0000} 91.08} \\
                          & SAM-Med2D~\cite{cheng2023sammed2d}  & 55.29 & 63.27       & 62.57        & 48.18       & 50.12    & 61.29  & 68.10   & 58.40                        \\
\multirow{-5}{*}{\begin{tabular}[c]{@{}c@{}}AMOS-\\ CT+MRI*\end{tabular}} & Ours & {\color[HTML]{3531FF} 94.38} & {\color[HTML]{FE0000} 89.20} & {\color[HTML]{3531FF} 88.16} & {\color[HTML]{3531FF} 76.37} & {\color[HTML]{3531FF} 69.12} & {\color[HTML]{3531FF} 91.23} & {\color[HTML]{3531FF} 85.34} & {\color[HTML]{3531FF} 84.83} \\ \hline
\end{tabular}%
}

\end{table}

\begin{table}[!t]
\centering
\caption{Results on different methods on ATLAS2023, we use one point to test SAM-Med2D on test sets. The DoDNet and MED3D are not robust when the new modality is introduced. The text is in {\color[HTML]{FE0000}red} and in {\color[HTML]{3531FF} blue} for the best and second-best results.}
\label{tab:atlas-result}
\begin{tabular}{c|ccc|ccc}
\hline
\multirow{2}{*}{Methods} & \multicolumn{3}{c|}{DSC (\%)} & \multicolumn{3}{c}{mIoU (\%)} \\ \cline{2-7} 
                         & Liver    & Tumor    & Mean    & Liver    & Tumor    & Mean    \\ \hline
DoDNet\cite{zhang2021dodnet}                   & 79,54    & 24.03    & 52.34   & 75.31    & 19.3     & 47.31   \\
MED3D~\cite{chen2019med3d}                   & 73.12    & 19.55    & 46.34   & 65.3     & 16.82    & 41.06   \\
SAM-Med2D~\cite{cheng2023sammed2d}                & 80.02        & 51.28        & {\color[HTML]{3531FF} 65.65}   & 77.34        & 48.88        & {\color[HTML]{3531FF} 63.11}       \\ 
Ours                     & 89.43    & 57.53    & {\color[HTML]{FE0000} 73.48}   & 86.34    & 55.32    & {\color[HTML]{FE0000} 70.83}   \\ \hline

\end{tabular}

\end{table}
%%%

\begin{table}[!t]
\centering
\caption{Ablation study on different model configurations.
Encoder or decoder marked with \textcolor{red}{\ding{55}} indicates the use of an MLP as the output layer for the Swin Transformer block. In contrast, \textcolor{green}{\ding{51}} denotes that the encoder or decoder incorporates our M\textsuperscript{4}oE structure. 
'GN' stands for gating network. 
The ablation studies reveal that employing the M\textsuperscript{4}oE architecture in both encoder and decoder significantly enhances model performance. 
The text is in {\color[HTML]{FE0000}red} and in {\color[HTML]{3531FF} blue} for the best and second-best results.}
\label{tab:abla_m4oe}
\resizebox{\textwidth}{!}{%
\begin{tabular}{cccc|ccccccccc}
\hline
\multicolumn{4}{c|}{Method}       & \multicolumn{9}{c}{DSC (\%)}                                          \\ \hline
Encoder &
  \multicolumn{1}{l}{Decoder} &
  \multicolumn{1}{l}{GN} &
  M\textsuperscript{4}oE &
  Liver &
  Liver Tumor &
  Left Kidney &
  Right Kidney &
  Gallbladder &
  Pancreas &
  Spleen &
  Stomach &
  Mean \\ \hline
\textcolor{red}{\ding{55}} & \textcolor{red}{\ding{55}} & \textcolor{red}{\ding{55}} & \textcolor{red}{\ding{55}} & 93.61 & 51.23 & 84.65 & 84.51 & 69.34 & 63.21 & {\color[HTML]{3531FF} 92.37} & 81.20 & 77.52 \\
\textcolor{green}{\ding{51}} & \textcolor{red}{\ding{55}} & \textcolor{green}{\ding{51}} & \textcolor{green}{\ding{51}} & 93.21 & 55.12 & 89.27 & 89.15 & 75.41 & 68.32 & 91.12 & 83.25 & 80.61 \\
\textcolor{green}{\ding{51}} & \textcolor{red}{\ding{55}} & \textcolor{red}{\ding{55}} & \textcolor{green}{\ding{51}} & 93.58 & 53.21 & 90.01 & {\color[HTML]{FE0000} 90.60} & {\color[HTML]{3531FF} 76.01} & 65.33 & 90.87 & 82.79 & 80.30 \\
\textcolor{red}{\ding{55}} & \textcolor{green}{\ding{51}} & \textcolor{green}{\ding{51}} & \textcolor{green}{\ding{51}} &  94.08 & 53.25 & 88.98 & 89.01 & 75.76 & {\color[HTML]{FE0000} 69.31} & {\color[HTML]{3531FF} 92.37} & {\color[HTML]{3531FF} 85.81} & {\color[HTML]{3531FF} 81.07} \\
\textcolor{red}{\ding{55}} & \textcolor{green}{\ding{51}} & \textcolor{red}{\ding{55}} & \textcolor{green}{\ding{51}} & {\color[HTML]{FE0000} 94.30}  & 53.19 & {\color[HTML]{3531FF} 90.12} & 88.91 & 74.11 & 68.45 & 91.36 & 85.43 & 80.73 \\
\textcolor{green}{\ding{51}} & \textcolor{green}{\ding{51}} & \textcolor{red}{\ding{55}} & \textcolor{green}{\ding{51}} & 93.12 & {\color[HTML]{3531FF} 55.37} & 88.67 & 87.56 & 75.39 & 66.29 & 90.12 & 83.04 & 79.95 \\
\textcolor{green}{\ding{51}} &
  \textcolor{green}{\ding{51}} &
  \textcolor{green}{\ding{51}} &
  \textcolor{green}{\ding{51}} &
  {\color[HTML]{3531FF} 94.12} &
  {\color[HTML]{FE0000} 57.53} &
  {\color[HTML]{FE0000} 90.85} &
  {\color[HTML]{3531FF} 90.46} &
  {\color[HTML]{FE0000} 76.56} &
  {\color[HTML]{3531FF} 69.16} &
  {\color[HTML]{FE0000} 92.52} &
  {\color[HTML]{FE0000} 85.97} &
  {\color[HTML]{FE0000} 82.15} \\ \hline
\end{tabular}%
}

\end{table}

\subsubsection{Ablation Study}
The ablation study in Table~\ref{tab:abla_m4oe} assesses the impact of implementing the M\textsuperscript{4}oE architecture in different components of a segmentation network. 
It indicates that integrating the M\textsuperscript{4}oE structure into both the encoder and decoder significantly enhances the model's segmentation performance, as evidenced by the highest mean DSC score of 82.15\%. 
Models using M\textsuperscript{4}oE in either the encoder or decoder showed improved results over the baseline model without M\textsuperscript{4}oE, but the full incorporation into both yielded the best outcomes.
The M\textsuperscript{4}oE architecture benefits the backbone model by enabling effective multimodal data integration, capturing diverse contextual information at multiple scales, and modulating features to emphasize relevant patterns for accurate segmentation. 
Additionally, the synergy between the encoder and decoder when both use the M\textsuperscript{4}oE structure results in more coherent feature learning and reconstruction. The presence of a gating network also contributes to performance improvements, suggesting its role in enhancing feature flow control within the network.

\section{Conclusion}
The proposed Medical Multimodal Mixture of Experts (M\textsuperscript{4}oE) presents a novel and efficient direction toward developing foundation models for multimodal medical image segmentation. 
By effectively integrating modality-specific experts within a flexible gating mechanism, M\textsuperscript{4}oE demonstrates accurate and adaptable segmentation across diverse imaging datasets.
We believe the expert specialization and dynamic ensemble concepts presented here will provide guiding principles for tackling multimodal challenges in medical imaging and beyond.
For future direction, evaluating M\textsuperscript{4}oE over larger-scale datasets, augmenting with uncertainty estimation, and enhancing model transparency could further facilitate clinical integration.

\bibliographystyle{splncs04}
\bibliography{5_ref.bib}

\end{document}